\documentclass[aps,prev,preprintnumbers,floatfix,nofootinbib,twocolumn]{revtex4-1}

\pdfoutput=1

\usepackage{mathrsfs, amsmath, amsthm, amssymb, color, epstopdf, verbatim, enumerate, graphicx, hyperref}

\begin{document}

\title{Are nuclear decay anomalies evidence of axion dark matter?}
\title{Nuclear decay anomalies as a signature of axion dark matter}
\author{Xin Zhang$\,{}^{a,b}$}
\author{Nick Houston$\,{}^{c\dagger}$}\email{nhouston@bjut.edu.cn}
\thanks{corresponding author}
\author{Tianjun Li$\,{}^{d,e\dagger}$}
\affiliation{	
        ${}^a$ National Astronomical Observatories, Chinese Academy of Sciences, 20A, Datun Road, Chaoyang District, Beijing 100101, China\\
		${}^b$ School of Astronomy and Space Science, University of Chinese Academy of Sciences, Beijing 100049, China\\
        ${}^c$ Institute of Theoretical Physics, Faculty of Science, Beijing University of Technology, Beijing 100124, China\\
		${}^d$ CAS Key Laboratory of Theoretical Physics, Institute of Theoretical Physics, Chinese Academy of Sciences, Beijing 100190, China\\
		${}^e$ School of Physical Sciences, University of Chinese Academy of Sciences, No. 19A Yuquan Road, Beijing 100049, China
		}

\begin{abstract}
 A number of nuclear decay anomalies have been reported in the literature, which purport to show periodic variations in the decay rates of certain radioisotopes.
 If these reports reflect reality, they would necessitate a seismic shift in our understanding of fundamental physics.
 We provide the first mechanism to explain these findings, via the misalignment mechanism of QCD axion dark matter, wherein oscillations of the effective $\theta$ angle induce periodic variation in nuclear binding energies and hence decay rates.
 As we expect this effect to be most pronounced in low-$Q$ systems, we analyse 12 years of tritium decay data ($Q\simeq$ 18.6 keV) taken at the European Commission's Joint Research Centre.
 Finding no statistically significant excess, we exclude axion decay constants below $9.4\times10^{12} - 1.8\times10^{10}$ GeV (95 \% confidence level) for masses in the $1.7\times{10}^{-23} - 8.7\times 10^{-21}$ eV range.
\end{abstract}

% keywords: axion, dark matter, tritium, decay anomaly

\maketitle 

\textbf{Introduction.} 
As a cornerstone of modern physics, it is widely accepted that radioactive decay is in general a truly random process, occurring independently of time, space and external influence.
This simple fact carries with it a vast array of consequences, across fields as diverse as modern-day nuclear medicine to the Big Bang Nucleosynthesis which occurred in the first few minutes of our cosmic history.

Nonetheless, in recent decades a number of purported nuclear decay anomalies have been reported in the literature, which represent apparent violations of this rule \cite{McDuffie:2020uuv, Fischbach:2009zz}.
Generally these anomalies take the form of periodic variations in the observed decay rate, although notably there are some indications of temporary variations in nuclear decay correlated with solar flares and other astrophysical phenomena \cite{Davis:1995wj,Jenkins:2008tt, Mohsinally:2016juj, Fischbach:2018dnd}.

One economical explanation of these conclusions, supported by a body of evidence, is that they are largely the result of random noise, unaccounted-for systematic effects and incomplete uncertainty analysis \cite{Pomme:2016yoh, Pomme:2016i, Pomme:2016ckl, Pomme:2016iii, Pomme:2018vgo, Breur:2019coe, Pomme:2020veo, Pomme:2019mhq, Pomme:2015}.
For example, the overall preference of some datasets for an annually modulated signal may be simply attributed to the influence of seasonally varying environmental conditions, rather than any exotic deviations from known physics \cite{Pomme:2019mhq}.
It should also be noted that although radioactive decay is of course a relatively well-explored and understood topic, accurately quantifying the various associated uncertainties is nonetheless nontrivial \cite{Pomme:2015}.

This perspective may be supported by the fact that such anomalies are furthermore somewhat difficult to explain in the context of ordinary particle physics.
Given the annual periodicity of some claimed signals, solar neutrinos are often suggested to be in some way responsible, but there is no known mechanism within the Standard Model which allows this \cite{McDuffie:2020uuv}.
Some somewhat speculative hypotheses have been proposed, as summarised in \cite{Pomme:2016ckl, Pomme:2022ead}, but at present there appears to be no concrete framework within which these anomalies can be understood consistently with other observations.

Nonetheless, the absence so far of a compelling explanatory framework does not mean that none exist to be found.
Furthermore, regardless of the provenance of these unusual observations, and in particular the key question of if they indeed represent a signature of new physics, if they can be connected to some favoured model of physics beyond the Standard Model, they can provide novel constraints and experimental strategies.

Bearing these motivations in mind we in the following provide the first mechanism to explain this phenomenon, via a class of particles known as axions.
Such an approach carries with it the benefit of also addressing certain other problems in fundamental physics, namely the nature of dark matter (DM) and the Strong CP problem in the Standard Model. 

Arising originally via a minimal extension of the Standard Model; the Peccei-Quinn solution of the Strong CP problem \cite{Peccei:1977ur, Weinberg:1977ma,Wilczek:1977pj}, axions and axion-like particles occupy a rare focal point in theoretical physics, in that they are simultaneously a generic prediction of the exotic physics of string and M-theory compactifications \cite{Svrcek:2006yi, Arvanitaki:2009fg,Visinelli:2018utg}.
Despite the profound differences between these two contexts the resulting axion properties are also largely universal, providing an easily-characterisable theoretical target.
As light, long lived pseudoscalar particles they can also influence many aspects of cosmology and astrophysics, creating a wide variety of observational signatures \cite{Marsh:2015xka}.
In particular, they are a natural candidate for the mysterious DM comprising much of the mass of our visible universe \cite{Dine:1982ah, Preskill:1982cy}, and as such are a topic of intense ongoing investigation \cite{Irastorza:2018dyq}.

The aforementioned Strong CP problem is the question of why the effective QCD $\theta$ parameter, which enters via the Lagrangian term
\begin{equation}
	\mathcal{L}_\theta = \frac{g^2\theta}{32\pi^2} G_{\mu\nu} \tilde G^{\mu\nu}\,,
\end{equation}
is smaller than $10^{-10}$ in absolute value and not $\mathcal{O}(1)$, as would be expected on the grounds of naturalness.
Here $g$ is the gauge coupling and $G_{\mu\nu}$ the gluon field strength tensor.
Axions which enact the Peccei-Quinn solution of the strong CP problem do so by promoting $\theta$ to a dynamical variable, which can then relax to zero as required.

In the event that these axions also comprise the DM in our Universe, the misalignment mechanism ensures that the time-dependent $\theta$ angle today is
\begin{equation}
	\label{eq: theta}
	\theta \simeq \sqrt{\frac{2\rho_{DM}}{m_a^2f_a^2}}\cos(\omega t +\vec p\cdot \vec x+\phi)\,,
\end{equation}
where $\rho_{DM}$ is the DM density, $\omega=m_a(1+\frac{1}{2}v^2+\mathcal{O}(v^4))$, $p$, $v$, $f_a$ and $m_a$ are the axion field momentum, velocity, decay constant and mass respectively, and $\phi$ is an arbitrary phase.

Since various aspects of nuclear physics are $\theta$-dependent this has the potential to lead to a variety of observational signatures, from which constraints on the axion parameter space can then be placed.
For example, Refs.~\cite{Abel:2017rtm, Roussy:2020ily, Aybas:2021nvn, Schulthess:2022pbp} search for an oscillating neutron electric dipole moment (nEDM), and in Ref.~\cite{Blum:2014vsa} it was also demonstrated that an oscillating $\theta$-angle can lead to underproduction of ${}^4$He during Big Bang Nucleosynthesis (BBN).

For our purposes it is the $\theta$-dependence of nuclear binding energy that is of primary interest.
Once nuclear binding energies become time-dependent via \eqref{eq: theta} we can expect a periodic variation in the nuclear decay rates and hence the possibility to explain the reported decay anomalies.

In particular, we will demonstrate that tritium represents a particularly opportune target to search for these effects.
This established, we will then use existing ${}^3$H data to search for the corresponding axion-induced signal.
Finding no statistically significant excess, we can then exclude axion decay constants below $9.4\times10^{12} - 1.8\times10^{10}$ GeV (95 \% confidence level) for masses in the $1.7\times10^{-23} - 8.7\times 10^{-21}$ eV range.
Conclusions and discussion are presented in closing.

\textbf{$\theta$-dependence of nuclear decay rates.} 
Our underlying quantity of interest is the fractional change in the beta decay rate $\Gamma(\theta)$ as a function of $\theta$,
\begin{equation}
    \label{eq: fractional change}
	I_0(\theta)\equiv\frac{\Gamma(\theta)-\Gamma(0)}{\Gamma(0)}\,,\,\,
	\Gamma(\theta)= \int_{m_e}^{E_{\rm max}+\delta E(\theta)}dE_e\,\frac{d\Gamma}{dE_e}\,,
\end{equation} 
where $E_{\rm max} = (M_i^2 +m_e^2 - (M_f + m_\nu)^2)/2M_i$, with initial and final state masses $M_{i/f}$, neutrino and electron masses $m_{\nu/e}$, $E_e$ the energy of the emitted electron, and $\delta E(\theta)$ an additional $\theta$-dependent contribution. 

In principle this $\theta$-dependence can also enter in other ways, for example modification of the underlying nuclear couplings $g_{A/V}$.
However, in line with the results of Ref.~\cite{Lee:2020tmi} it is reasonable to assume the leading order contributions arise specifically from modifications of the phase space.
This point is also specifically analysed in the Supplementary Material~\cite{supplementary material}, in support of this conclusion.

By approximating nuclear beta decay in terms of free neutron decay, it has also been argued in Ref.~\cite{Fischbach:2009zz} that for small perturbations to the decay rate, decays with smaller $Q \equiv M_i-M_f - m_e$ resulted in a larger fractional change in the beta decay rate.

However, from the beta decay rates given in Ref.~\cite{Faessler:2011zz, Dvornicky:2011fm, Doi:1985dx} we can in any case evaluate this integral numerically for small $\delta E(\theta)$ without the free-neutron approximation, finding for ${}^{187}$Re ($Q\simeq2.6$ keV) and ${}^3$H ($Q\simeq18.6$ keV) that
\begin{equation}
	I_0(\theta) \big|_{{}^{187}Re}\simeq 1.16\left(\frac{\delta E(\theta)}{\mathrm{keV}}\right)\,,\quad
	I_0(\theta) \big|_{{}^{3}H}\simeq 0.18\left(\frac{\delta E(\theta)}{\mathrm{keV}}\right)\,.
	\label{variation}
\end{equation}

This result is particularly fortuitous in that if we wish to search for this effect, high quality datasets for ${}^3$H already exist.
The use of ${}^3$H decay as a window to new physics more generally has also recently been explored in Ref.~\cite{Canning:2022nye,Dror:2019onn,Dror:2019dib}.
Furthermore, in addition to being more sensitive to the effect of a time-varying $\theta$ angle, the $\theta$-dependence of the properties of lighter nuclei such as ${}^3$H may also be comparatively easier to discern.

Indeed, in Ref.~\cite{Lee:2020tmi} the dependence of the nuclear binding energy upon $\theta$ has already been calculated for light nuclei.
It was estimated that for three and four-nucleon systems the $n$-nucleon binding energy $\overline B_n$ satisfies
\begin{equation}
	\label{light nuclei binding energy}
	\overline B_n(\theta)^{1/4}-\overline B_2(\theta)^{1/4}
	=\overline B_n(0)^{1/4}-\overline B_2(0)^{1/4}\,.
\end{equation}
Here the bar indicates an average over states which become degenerate in the limit that the approximate Wigner $SU(4)$ symmetry of low energy nuclear physics becomes exact.
As noted in Ref.~\cite{Lee:2020tmi} this agrees with the results calculated numerically in Ref.~\cite{Barnea:2013uqa}.

The averaged 2-nucleon binding energy $\overline{B_2}(\theta)$ can be found by averaging over the physical deuteron and the spin singlet (dineutron and diproton) channel, with $\theta$ dependence parameterised via
\begin{equation}
	B_2(\theta) = \left(B_2(0) +\sum_{i=1}^3c_i(1-\cos\theta)^i\right)\,\,\text{MeV}\,,
\end{equation}
where $B_2(0) = 2.22, -0.072, -0.787$ MeV for the deuteron, dineutron and diproton respectively, and the $c_i$ are numerical coefficients given in Ref.~\cite{Lee:2020tmi} (herein we choose the more conservative parameter set II, assuming isospin conservation).

Since this only gives the (averaged) $\overline B_3(\theta)$, rather than the individual initial and final state binding energies $B(\theta)_{i/f}$, we assume in the following that for small $\theta$ 
\begin{equation}
    \label{eq: binding energy difference}
	B(\theta)_{i/f}
    \simeq B(0)_{i/f}\frac{\overline B_3(\theta)}{\overline B_3(0)}\,,
\end{equation}
which is equivalent to assuming that for small $\theta$ the individual binding energies scale in proportion to their average.

Knowing the $\theta$-dependence of the binding energy, we can then calculate $\delta M_{i/f}$, the $\theta$-dependent part of $M_{i/f}$, via
\begin{equation}
    M_{i/f} = \sum_N m_N(\theta) - B(\theta)_{i/f}\,,
\end{equation}
where the sum runs over nucleons, and the neutron/proton mass difference is given by
\begin{equation}
    \label{eq: neutron proton mass difference}
    m_n - m_p
    \simeq -0.58{\rm MeV}+4c_5B_0\frac{m_\pi^2}{m_\pi^2(\theta)}\left(m_u - m_d\right)\,,
\end{equation}
where $c_5=(-0.074\pm0.006)$GeV${}^{-1}$ is a low energy constant, $B_0 = m_\pi^2/(m_u+m_d)$, and we assume two degenerate quark flavours, following Ref.~\cite{Lee:2020tmi, Gasser:2020mzy}

Expanding about $\theta =0$ we find 
\begin{equation}
    \delta E(\theta) \simeq \delta M_{i} - \delta M_f\,,
\end{equation}
which in turn can be calculated via equations \eqref{eq: binding energy difference} and \eqref{eq: neutron proton mass difference} since we know the $\theta$-dependent parts of $(m_n - m_p)$ and $(B_i - B_f)$.
In turn this allows us to calculate $I_0(\theta)$.

However, care is required here.
In a Universe where this mechanism is active the measured values of physical quantities such as decay rates are actually not those at $\theta=0$, but are instead those at $\langle\theta^2\rangle$ (since binding energies are $\mathcal{O}(\theta^2)$ at leading order).
Therefore when dealing with experimental data, rather than comparing $\Gamma(\theta)$ to $\Gamma(0)$, we should instead consider
\begin{equation}
    I(\theta)
    \equiv\frac{\Gamma(\theta)-\langle\Gamma\rangle}{\langle\Gamma\rangle}
    =\frac{\Gamma(\theta)}{\Gamma(0)}
    \left(\frac{\Gamma(0)}{\langle\Gamma\rangle}\right) - 1\,,
\end{equation}
where $\langle\Gamma\rangle$ is the average value of $\Gamma(\theta)$.
Using the known $I_0(\theta) = \Gamma(\theta)/\Gamma(0)-1$, we find after some calculation
\begin{align}
    I(\theta) 
    \simeq &\,1.3\times10^{-5}\left(\frac{\rho_{DM}}{0.45\rm{GeV}/\mathrm{cm}^3}\right) \left(\frac{10^{-21}\rm{eV}}{m_a}\right)^2\nonumber\\
    &\left(\frac{10^{13}\rm{GeV}}{f_a}\right)^2
    \cos(2(\omega t +\vec p\cdot \vec x+\phi))\,,
\end{align}
which oscillates about zero, as expected, leading to alternating excesses and deficits in the number of nuclear decays per unit time.
We can also see that the corresponding shift in the decay energy,
\begin{align}
    |\delta E(\theta)|
    \simeq  &\, 3.5\times10^{-2}{\rm eV}\left(\frac{\rho_{DM}}{0.45\rm{GeV}/\mathrm{cm}^3}\right) \left(\frac{10^{-21}\rm{eV}}{m_a}\right)^2\nonumber\\
    &\left(\frac{10^{13}\rm{GeV}}{f_a}\right)^2
\end{align}
is also a small correction in regions of the axion parameter space we exclude, justifying our assumptions that $\delta E<< Q$.
Similarly, $\theta<<1$ in these regions.

\textbf{Data analysis and results.} 
These points established, we can now search for evidence of this effect in ${}^3$H.
We make use of a dataset, shown in Fig.~\ref{fig: time-series data}, provided by the European Commission's Joint Research Centre (JRC) at the Directorate for Nuclear Safety and Security in Belgium, which spans approximately 12 years of liquid scintillation counter observations of the decay of an $\mathcal{O}({\rm microcurie})$ ${}^3$H source \cite{Pomme:2016ckl}.
\begin{figure}[h]
	\centering
	\includegraphics[width=1\columnwidth]{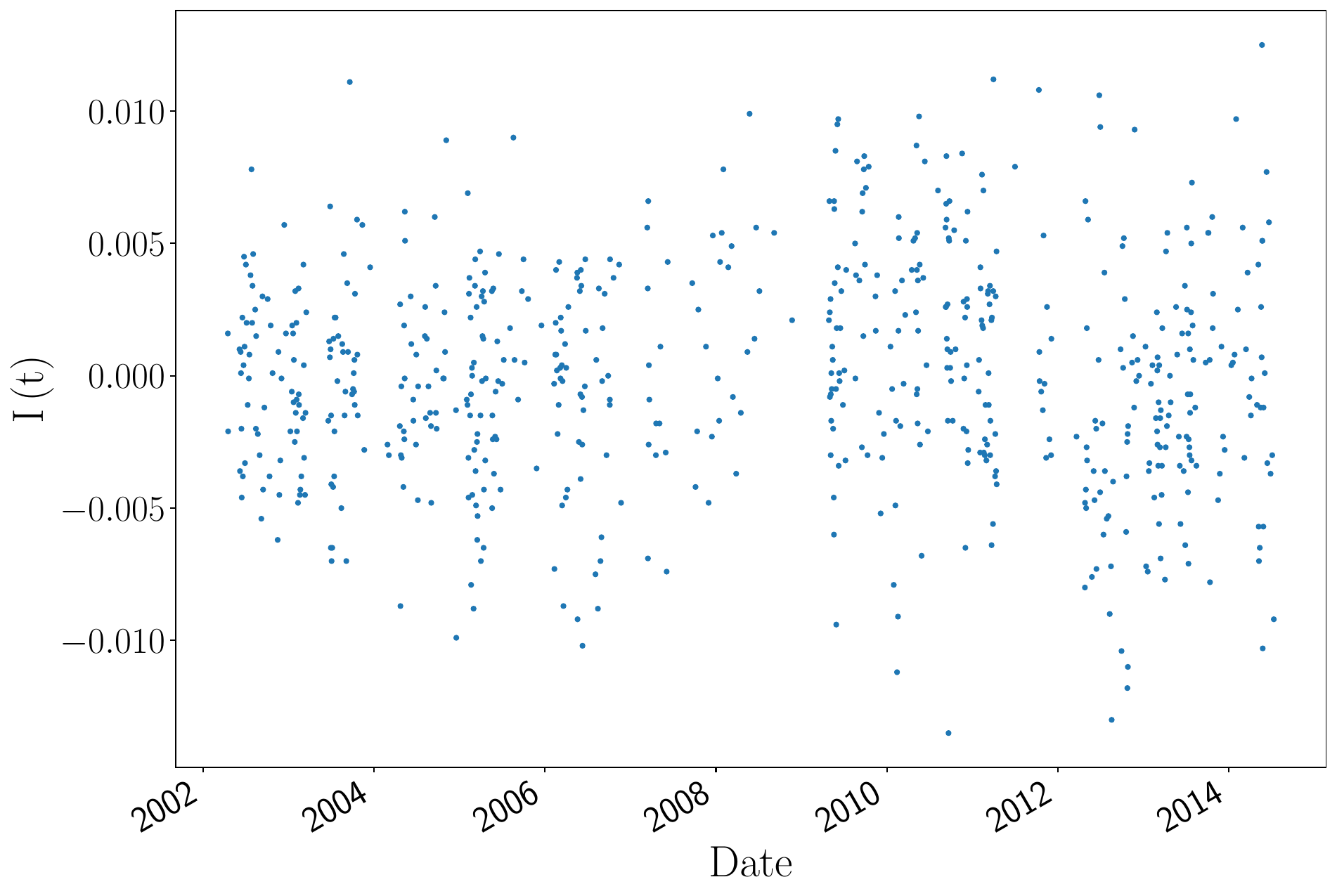}
	\caption{Time-series data from Ref.~\cite{Pomme:2016ckl}, showing the fractional change in the ${}^3$H beta decay rate, $I(t) = (\dot N - \langle\dot N\rangle)/\langle\dot N\rangle$. 
    Here $\dot N$ is the measured value of decays per second, whilst $\langle\dot N\rangle$ is its expected value due to the exponential decay law.
    }
	\label{fig: time-series data}
\end{figure}

It is worth emphasising that the statistical analysis of nuclear decay anomalies has been subject to a number of treatments \cite{Pomme:2020noq}, with no overall consensus on which approach is optimal.
This being the case we broadly follow the approach of Refs.~\cite{Abel:2017rtm, Wu:2019exd, Aybas:2021nvn}, where searches for oscillatory signals in time-series data were also used to place limits on ultralight axion DM.

Since the data points are unevenly spaced in time we will estimate their power spectrum using the Least Squares Spectral Analysis (LSSA) method to construct periodograms \cite{Lomb:1976wy, Scargle:1982bw}. 
We compute the power spectrum using the {\tt astropy.timeseries.LombScargle} class provided by the {\tt Python astropy.timeseries} package\cite{VanderPlas:2018bw}, evaluated at a set of $8113$ evenly spaced trial frequencies. 

As the lowest $15$ trial frequencies appear to show evidence of uncontrolled systematic effects, possibly long-term drift of the experimental apparatus, we exclude them from our analysis.
The largest frequency in the remaining data, $4.2\times10^{-6}$ Hz, corresponds to a period of $\sim2.8$ days, whilst the smallest, $8.0\times10^{-9}$ Hz, corresponds to a period of $\sim4.0$ years. 
The resulting periodogram is then given by the blue curve in Fig.~\ref{fig: periodogram}.

Under the null hypothesis that the dataset contains no axion-induced signal, the time-series datapoints should follow a Gaussian distribution about $I = 0$. 
To place limits on the corresponding power spectrum we therefore perform Monte Carlo (MC) simulations by generating $N=50000$ time-series MC datasets, with the same time spacing as the original data.
The MC data points themselves are drawn from a zero-mean Gaussian distribution, with width set by the standard deviation of the original dataset.

For each MC dataset we can calculate a corresponding periodogram, and then use the statistics of these periodograms to construct the Cumulative Distribution Function (CDF) for the power at each frequency.
From these CDFs we then determine the false positive (or false alarm) power at 95\% confidence level for each frequency, as shown in Fig.~\ref{fig: periodogram} in orange.
Here and in the following we account for the `look elsewhere effect' by defining these limits with respect to the global trials factor $p_{\rm global} = 1-(1 - p_{\rm local})^{N_f}$, where $p_{\rm local}$ is the corresponding local $p$-value and $N_f$ the number of trial frequencies.
As can be seen, nowhere does the original dataset exceed the 95 \% confidence level threshold, indicating the data are compatible with the null hypothesis.

\begin{figure}[h]
	\centering
	\includegraphics[width=1\columnwidth]{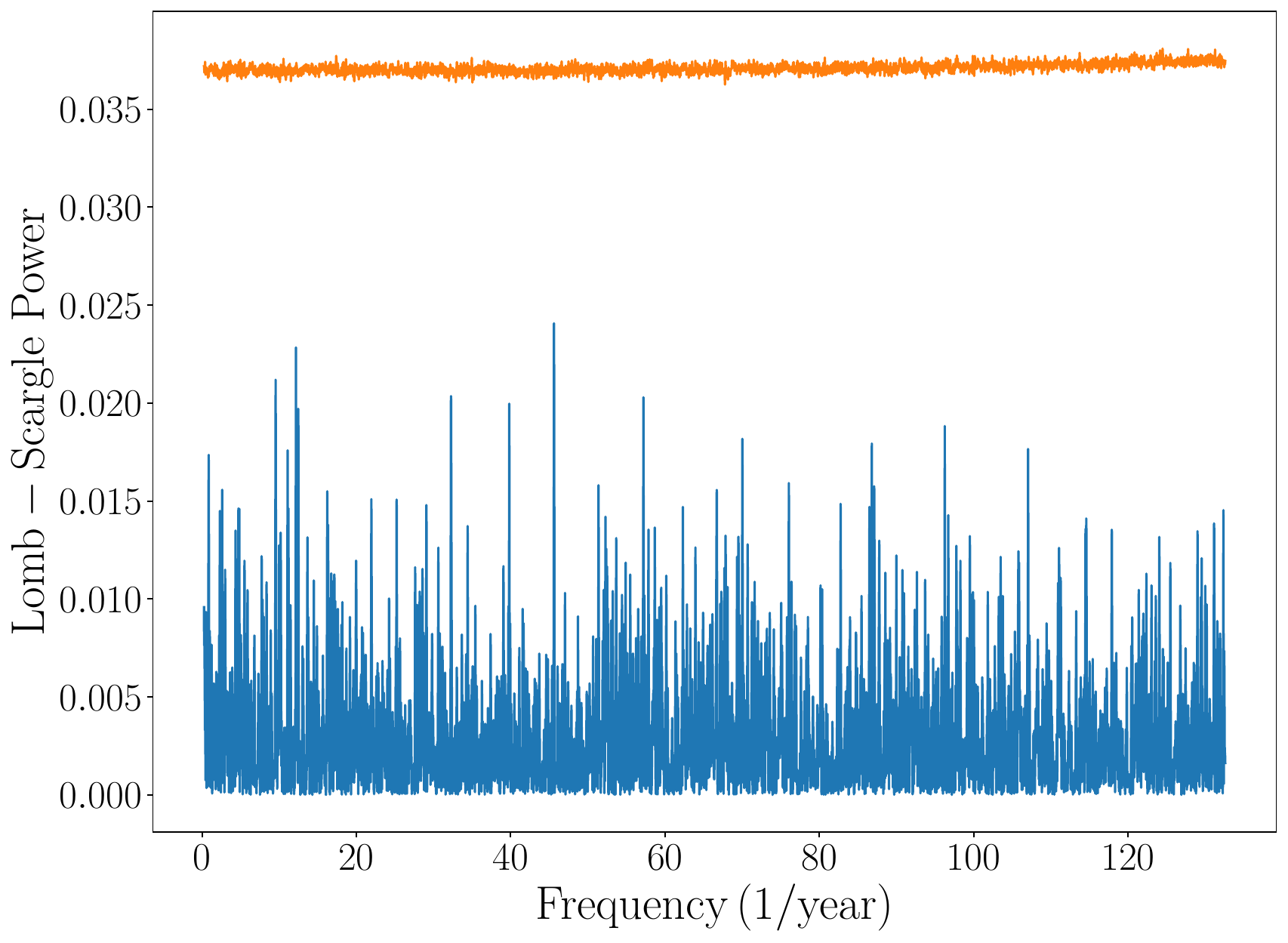}
	\caption{Periodogram corresponding to the dataset shown in Fig.~\ref{fig: time-series data} (blue), along with the MC-derived 95\% confidence level threshold (orange).
    As can be easily seen, the data are compatible with the null hypothesis.
	}
	\label{fig: periodogram}
\end{figure}

To determine the corresponding limit on the axion parameters we follow a similar approach.
For each trial frequency we construct $N=50000$ MC datasets containing Gaussian background and injected axion-derived signals and calculate their periodograms.
The injected signals are of fixed amplitude and frequency, with the unknown axion phase $\phi$ drawn from a uniform distribution.
As the injected signals are constructed in time series form, we directly match the axion velocity to the lab-frame DM velocity, incorporating modulation effects due to the Earth's passage through the DM halo \cite{Bandyopadhyay:2010zj, Freese:2012xd}.
For a given choice of parameters, we can then construct the corresponding CDF for the power at each trial frequency.
With the mass fixed by the frequency under consideration, the threshold value of $f_a$ can then be determined following a standard frequentist approach, well illustrated in Ref.~\cite{Centers:2019dyn}.
Having determined the threshold value of $f_a$ within this framework, we then subsequently account for the `stochastic vs deterministic' correction factor occurring in the regime where the measurement time is much less than axion field coherence time, following Ref.~\cite{Centers:2019dyn}.

Specifically, from the background-only CDF we first find the false positive threshold power at a desired confidence level.
From the background plus signal CDF we can also find the false negative threshold power, for a given choice of $f_a$.
The threshold value of $f_a$ is then determined by the condition that the false positive threshold from the background-only CDF coincides with the false negative threshold from the background plus signal CDF, at the desired confidence level.
Equivalently we can say that the threshold value of $f_a$ occurs when the false positive rate $\alpha$ is equal to the false negative rate $1-\beta$, where $CL=1-\alpha$.
The resulting exclusion curve is given in Fig.~\ref{fig: final constraint}.

\begin{figure}[h]
	\centering
	\includegraphics[width=1\columnwidth]{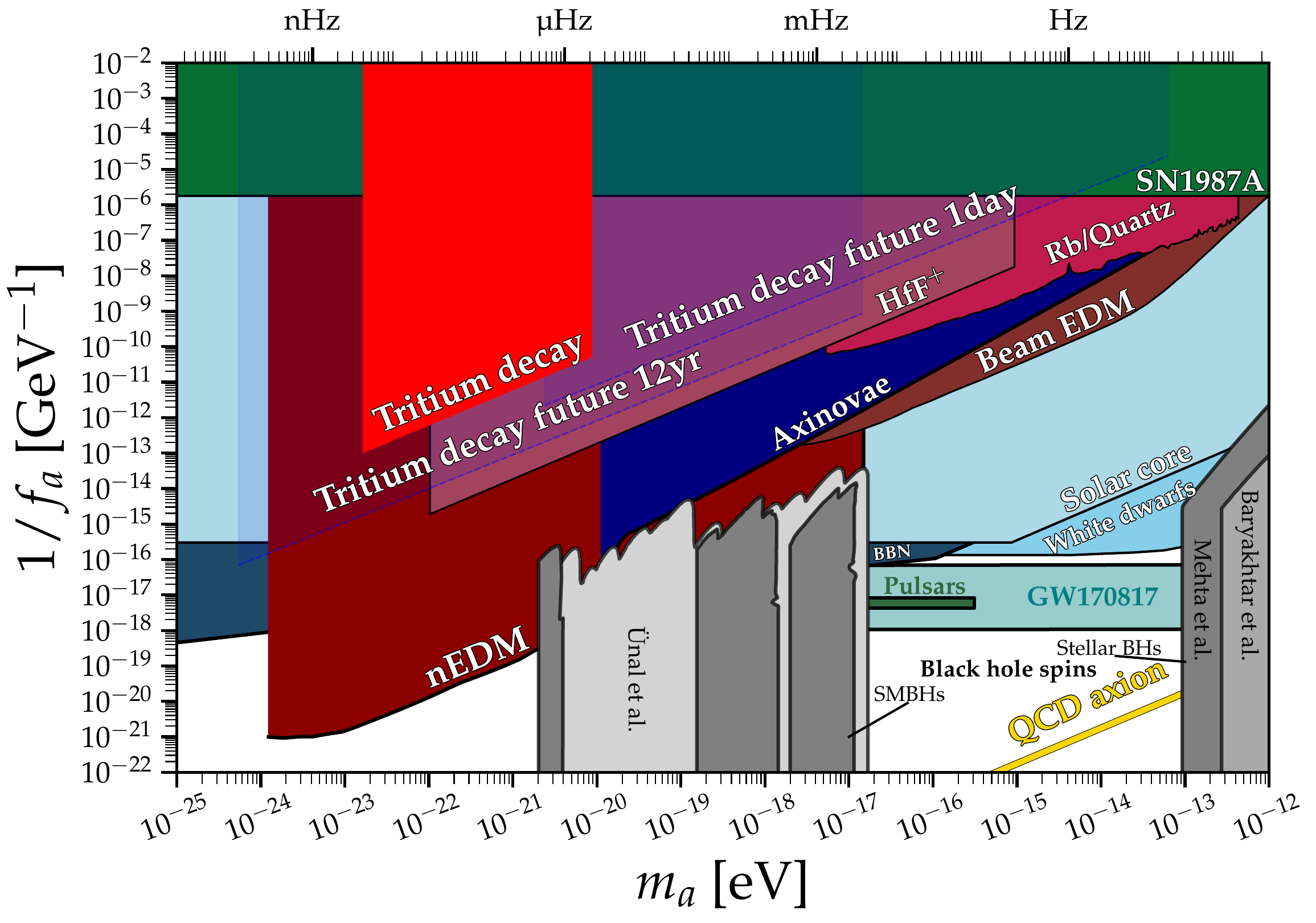}
	\caption{Axion DM constraint from the non-observation of periodic variations in ${}^3$H decay data.
	We exclude $f_a$ below $9.4\times10^{12} - 1.8\times10^{10}$ GeV (95 \% confidence level) for masses in the $1.7\times10^{-23} - 8.7\times 10^{-21}$ eV range.
	Also shown are constraints from oscillating nEDM searches \cite{Abel:2017rtm, Roussy:2020ily, Aybas:2021nvn, Schulthess:2022pbp}, BBN \cite{Blum:2014vsa}, the spectroscopy of radio-frequency atomic transitions \cite{Zhang:2022ewz}, pulsars, gravitational waves and black hole superradiance \cite{Hook:2017psm, Zhang:2021mks, Baryakhtar:2020gao, Unal:2020jiy, Stott:2020gjj}, solar/white dwarf observations \cite{DiLuzio:2021gos, Balkin:2022qer}, so-called `axinovae' \cite{Fox:2023aat}, and the parameter space occupied by canonical QCD axion models (yellow).
	Figure producing using the AxionLimits code \cite{AxionLimits}.
}
	\label{fig: final constraint}
\end{figure}

We simulate future experimental possibilities, via two fiducial cases also shown in Fig.~\ref{fig: final constraint}.
The first is a long-term experiment designed to search lower frequencies, with 1 measurement per hour over 12 years. 
The second is a short-term experiment to search higher frequencies, with approximately 1 measurement per second over 1 day. 
Both schemes increase the initial quantity of tritium by a factor of 100 relative to the JRC dataset, up to a presumed limit imposed by detector pileup \cite{Luo:2018zmx}.
In the first case we can exclude axion decay constants up to $f_a<1.5\times10^{16}$ GeV (95 \% confidence level) and cover masses in the $5.4\times10^{-25} - 1.4\times 10^{-17}$ eV range, while in the latter we can exclude axion decay constants up to $f_a<4.5\times10^{11}$ GeV (95 \% confidence level) and cover masses in the $2.4\times10^{-21} - 6.3\times 10^{-14}$ eV range.

\textbf{Conclusions/discussion.}
We have explored purported variations in nuclear decay rates, providing a novel and well-motivated explanatory mechanism within the framework of axion physics.
The oscillating QCD $\theta$-angle created by the misalignment mechanism induces a time-varying nuclear binding energy, which then leads to the time-dependence of nuclear decays.

Analysing 12 years of ${}^3$H decay data we find no corresponding statistically significant excess, and therefore exclude axion decay constants below $9.4\times10^{12} - 1.8\times10^{10}$ GeV (95 \% confidence level) for masses in the $1.7\times10^{-23} - 8.7\times 10^{-21}$ eV range.

One advantage of this approach is that we can probe regions of axion parameter space which normally are only accessible experimentally via much more sophisticated methods, such as oscillating nEDM searches \cite{Abel:2017rtm, Roussy:2020ily,Aybas:2021nvn, Schulthess:2022pbp}.
In contrast, the data we have analysed here were taken using a single $\mathcal{O}({\rm microcurie})$ tritium source and a commercially available laboratory liquid scintillation counter.  
Of course, there are also astrophysical and cosmological bounds on the mass of axion dark matter, such as $m_a>3\times10^{-19}$ eV (99 \% confidence level), coming from observations of ultra-faint dwarf galaxies \cite{Dalal:2022rmp} and $m_a<2\times10^{-20}$ eV (95 \% confidence level)bounds from the Lyman-alpha forest \cite{Rogers:2020ltq}. 
However, our work provides a new kind of laboratory-based scheme to search for QCD axion dark matter, which is complementary to such constraints without relying on cosmological or astrophysical assumptions and modelling.

However, relying on nuclear decays in this way also presents certain challenges. 
%In particular, we as a result operate in the signal-dominated regime, rather than the background-dominated regime more typical of axion DM searches.
In particular, issues such as detector pileup may ultimately constrain sensitivity more so than in other experimental approaches.

Another overall consideration here is the validity of this and other approaches in searching for QCD axion-derived phenomena far away from the parameter space where these effects are expected to occur: the so-called QCD axion band. 
It is first of all important to note that (admittedly non-minimal) models of the QCD axion do exist in the parameter space we have probed here, as given in Refs.~\cite{Hook:2018jle, DiLuzio:2021pxd, DiLuzio:2021gos, Banerjee:2022wzk}.
Furthermore, we can also understand our efforts here as being a necessary step to ultimately reach the sensitivity to probe standard QCD axion models via methods such as these.

There are several avenues by which these findings could be improved.
On the theoretical side, more accurate estimation of the dependence of nuclear binding energies on $\theta$ would be of use.
Experimentally, dedicated nuclear decay experiments which are optimised to search for this particular effect should also yield stronger constraints, and may be able to probe previously unexplored regions of the axion parameter space.
Analysis of other preexisting nuclear decay datasets may also similarly bear fruit.

\begin{acknowledgments} 
This work is supported in part by the National Key Research and Development Program of China Grant No. 2020YFC2201504, by the Projects No. 11875062, No. 11947302, No. 12047503, No. 12150410317 and No. 12275333 supported by the National Natural Science Foundation of China, by the Key Research Program of the Chinese Academy of Sciences, Grant NO. XDPB15, by the Scientific Instrument Developing Project of the Chinese Academy of Sciences, Grant No. YJKYYQ20190049, and by the International Partnership Program of Chinese Academy of Sciences for Grand Challenges, Grant No. 112311KYSB20210012. 
The data were taken by Dr. T. Altzitzoglou and generously provided by Dr Stefaan Pomm\'e, at the European Commission's Joint Research Centre at the Directorate for Nuclear Safety and Security in Belgium.
\end{acknowledgments}

\clearpage
\newpage
\maketitle
\onecolumngrid

\begin{center}
\textbf{\large Nuclear decay anomalies as a signature of axion dark matter} \\ 
\vspace{0.05in}
{ \it \large Supplementary Material}\\ 
\vspace{0.05in}
{}
{Xin Zhang, Nick Houston, Tianjun Li}

\end{center}

In this supplementary material we derive in detail the $\theta$-dependence of the nuclear half life of ${}^3$H.
This radioisotope is of particular interest due to the low $Q$-value ($18.6$ keV), which may result in greater sensitivity to the extremely small axion-induced energy shifts we are searching for. 
Conveniently ${}^3$H beta decay has for this reason also been studied intensively, due to the relevance it may have for cosmic neutrino background searches.

Our starting point is the Hamiltonian density 
\begin{equation}
	{\cal H}(x)=\frac{G_{\beta}}{\sqrt{2}} \bar{e} (x)\gamma^{\mu} (1- \gamma_5) \nu_{e}(x)j_\mu (x) + {h.c.},
\end{equation}
where $G_\beta = G_F \cos{\theta_C}$, $G_F$ is the Fermi constant, $\theta_C$ is the Cabbibo angle, and $e/\nu_{{e }}$ are respectively the electron and neutrino fields. 
The strangeness-conserving free nucleon charged current is
\begin{equation}
	j^{\mu} (x)=\bar{p}(x)\gamma^{\mu} (g_{{V}}-g_{{A}} \gamma_{{5}}) n(x)\,,
\end{equation}
where $p/n$ are the proton and neutron fields, and $g_V/g_A$ are the nuclear vector and axial-vector couplings respectively.
Following the detailed calculations in Ref.~\cite{Faessler:2011zz}, the total $\beta$-decay rate is
\begin{equation}
	\label{Eq: beta decay rate}
	\Gamma = \frac{1}{2\pi^3} m_e \left( G_\beta m_e^2 \right)^2\left( B_F({^{3}H}) + B_{GT} ({^{3}H}) \right)~I^\beta ({^{3}H})\,,
\end{equation}  
where the Fermi and Gamow-Teller beta strengths are respectively
\begin{eqnarray}
	&B_F({^{3}H})
	=g_V^2~|M_{F}|^{2} 
	=g_V^2~ \frac{1}{2}\left| {_{{^3}He}\langle }(1/2)^+ \parallel  \sum_n \tau^+_n (1/2)^+ {\rangle _{{^3}H}} \right|^2\,,\nonumber\\
	&B_{GT}({^{3}H}) 
	=g^2_A~|M_{GT}|^{2} 
	= g^2_A ~\frac{1}{2}\left| {_{{^3}He}\langle }(1/2)^+ \parallel  \sum_n \tau^+_n \sigma_n(1/2)^+{\rangle _{{^3}H}} \right|^2\,,
\end{eqnarray}
and the phase space integral is
\begin{equation}
	\label{Eq: phase space integral}
	I^\beta = \frac{1}{m_e^5}\int_{m_e}^{E_{\rm max}} F_0(Z+1,E_e) p_e E_e (E_{\rm max}-E_e)^2 dE_e\,,
\end{equation}
where $E_{\rm max}$ is the maximum possible electron energy, $p_e=|\vec p_e|$ is the magnitude of the electron 3-momentum, and $F_0$ is the Fermi function accounting for modification of electron wavefunctions due to the Coulomb field of the nucleus.

We will in the following assume isospin conservation, since isospin-violating effects and the $\theta$-dependence we seek to understand are both individually small corrections in our regime of interest.
These points established, we can assess where $\theta$-dependence may conceivably enter.

\noindent{\bf Non-QCD parameters.}
It is evident first of all that parameters which belong to non-QCD sectors, such as $m_e$, $G_F$, and $\theta_C$ are $\theta$-independent.
The PMNS matrix elements responsible for neutrino oscillations and the neutrino masses themselves should also presumably have no such dependence.

\noindent{\bf Nucleon masses.}
Conversely, it is already known that for free nucleons $(m_n - m_p)$ increases with increasing $\theta$,
\begin{equation}
	\label{Eq: neutron proton mass difference}
	m_n - m_p = \left(m_n-m_p\right)^{\rm{QED}}+\left(m_n-m_p\right)^{\rm{QCD}}\,,
\end{equation}
where $(m_n-m_p)^{\rm{QED}}=-(0.58 \pm 0.16) \rm{MeV}$ is the QED correction to the mass difference \cite{Gasser:2020mzy}.
We can then write
\begin{align}
    m_n - m_p &\simeq -0.58{\rm MeV}+4c_5B_0\frac{M_\pi^2}{M_\pi^2(\theta)}\left(m_u - m_d\right) \nonumber \\
    &\simeq (-0.58+1.87+0.21\theta^2){\rm MeV} \nonumber \\
    &\simeq (1.29+0.21\theta^2){\rm MeV}\,,
\end{align}
where $c_5=(-0.074\pm0.006)$GeV${}^{-1}$ is a low energy constant, $B_0 = m_\pi^2/(m_u+m_d)$, and we assume two degenerate quark flavours, following Ref.~\cite{Lee:2020tmi}.

\noindent{\bf Nuclear binding energies.}
We can also expect nuclear binding energies to be $\theta$-dependent.
To calculate this dependence we can make use of the results in Ref.~\cite{Lee:2020tmi}, where it was empirically estimated that for $n=3,4$ nucleon systems the binding energies $B_n$ satisfy
\begin{equation}
	(\overline B_n(\theta)/\overline B_4(0))^{1/4}-(\overline B_2(\theta)/\overline B_4(0))^{1/4}
	=(\overline B_n(0)/\overline B_4(0))^{1/4}-(\overline B_2(0)/\overline B_4(0))^{1/4}\,,
\end{equation}
Here the bar indicates an average over states which become degenerate in the limit that the approximate Wigner $SU(4)$ symmetry of low energy nuclear physics becomes exact.

Rearranging for $n=3$ yields
\begin{equation}
    \overline B_3(\theta)=\left(\overline B_2(\theta)^{1/4}+\overline B_3(0)^{1/4}-\overline B_2(0)^{1/4}\right)^4,
\end{equation}
where averaging over the physical ${}^3$H and ${}^3$He states gives $\overline B_3(0)\simeq8.1\,\mathrm{MeV}$, and $B_4(0)\simeq28.3\,\mathrm{MeV}$ is given by the physical ${}^4$He binding energy \cite{Lee:2020tmi}.

To find $\overline B_2(\theta)$, we average over the physical deuteron and the spin singlet (dineutron and diproton) channel, with $\theta$ dependence parameterised via 
\begin{equation}
	B_2(\theta) = \left(B_2(0) +\sum_{i=1}^3c_i(1-\cos\theta)^i\right)\,\,\text{MeV}\,,
\end{equation}
with $B_2(0) \simeq 2.22, -0.072, -0.787\, \text{MeV}$ for the deuteron, dineutron and diproton respectively, following Ref.~\cite{Lee:2020tmi}, which then yields $\overline B_2(0) \simeq 1.03\,\mathrm{MeV}$.

For the $c_i$ numerical coefficients we use the (more conservative) choice of parameter set II in Ref.~\cite{Lee:2020tmi}, assuming isospin conservation, ($c_1=3.25$, $c_2 = 2.55$, $c_3 = 0.47$).
Inserting these values we find
\begin{equation}
    \overline B_3(\theta) \simeq (8.1+7.63\,\theta^2){\rm MeV}\,,
\end{equation}
Since we only know the (averaged) $\overline B_3(\theta)$, rather than the individual initial and final state binding energies $B_i(\theta)$ and $ B_f(\theta)$, we will assume in the following that for small $\theta$
\begin{equation}
	B(\theta)_{i/f}
    \simeq B(0)_{i/f}\frac{\overline B_3(\theta)}{\overline B_3(0)}\,,
\end{equation}
which is equivalent to assuming that for small $\theta$ the individual binding energies scale in proportion to their average.

\noindent{\bf Initial/final state energies.}
Knowing the dependence of free nucleon masses and nuclear binding energies then allows us to calculate $M_i$ and $M_f$, the masses of the initial and final nuclear states. 
This is defined in terms of the sum of the masses of their constituent (free) nucleons $N$ and their associated binding energies,
\begin{equation}
	M_{i/f}(\theta) = \sum_N m_N(\theta) - B_{i/f}(\theta)\,,
\end{equation}
Combining the previous results we then find
\begin{equation}
    M_{i}(\theta) - M_{f}(\theta) 
    = (m_n - m_p)(\theta) - B_i(\theta) + B_f(\theta)
    \simeq 0.53-0.51\,\theta^2{\rm MeV}\,,
\end{equation}

\noindent{\bf Integral endpoint.}
Modifications to $M_{i/f}$ enter into the beta decay rate via the phase space integral in Eq.~\eqref{Eq: phase space integral}, through the upper limit $E_{\rm max}$, where
\begin{equation}
	E_{\rm max} = \frac{M_i^2 + m_e^2 - (M_f +m_\nu)^2}{2M_i}\,,
\end{equation}
where $m_\nu$ is the neutrino mass.
To calculate the leading order modification to $E_{\rm max}$ we can perturb $M_{i/f}$ by $\delta M_{i/f}$ (representing a $\theta$-dependent correction) and expand to lowest order, finding
\begin{equation}
	E_{\rm max} \simeq E_{\rm max}\big|_{\delta M_{i/f} = 0} 
    +\delta M_i \frac{M_i^2 - m_e^2 + (M_f +m_\nu)^2}{2M_i^2} 
    -\delta M_f \frac{M_f+m_\nu}{M_i}\,,
\end{equation}
Since $M_{i/f}>> m_e,m_\nu$ we can further neglect terms proportional to products of $\delta M_{i/f}$ and $m_{e}/m_\nu$ to then find
\begin{equation}
    E_{\rm max} \simeq E_{\rm max}\big|_{\delta M_{i/f} = 0} 
    +\delta M_i \frac{M_i^2 + M_f^2}{2M_i^2} 
    -\delta M_f \frac{M_f}{M_i}\,,
\end{equation}
The latter terms can be calculated exactly here, but we also can notice that since $Q = M_i - M_f - m_e\simeq18.6$ keV is also much less than $M_{i/f}$, we then have
\begin{align}
    \delta M_{i/f} M_f
    &= \delta M_{i/f} (M_i - m_e - Q) \nonumber \\
    &\simeq \delta M_{i/f} M_i\,,
\end{align}
which then leads simply to
\begin{equation}
    E_{\rm max} \simeq E_{\rm max}\big|_{\delta M_{i/f} = 0} 
    +\delta M_i -\delta M_f\,,
\end{equation}
From the previous section we can then calculate:
\begin{align}
    \label{eq: delta theta}
	E_{\rm max}(\theta) &\simeq E_{\rm max}(0)+\delta E(\theta) \nonumber \\
    &\simeq 0.53-0.51\,\theta^2 {\rm MeV}\,,
\end{align}

\noindent{\bf Fermi function.}
Before calculating the phase space integral we need to examine the relativistic Fermi function\cite{Faessler:2011zz, Dvornicky:2011fm, Doi:1985dx}, given by
\begin{equation}
	\label{eq: Fermi function}
	F_{k-1}(Z, E_e) =\left(\frac{\Gamma(2k+1)}{\Gamma(k)\Gamma(1+2\gamma_k)}\right)^2(2p_e R_0)^{2(\gamma_k-k)}|\Gamma(\gamma_k+iz)|^2e^{\pi z} \,,
\end{equation}
where $k = 1,2,3\dots$, $Z$ is the atomic number, $\gamma_k=\sqrt{k^2-(\alpha Z)^2}$, $R_0 = 1.2\times 10^{-15}$m is the nuclear radius, and $z = \alpha Z E_e/p_e$.
In principle $\theta$-dependence may enter via $R_0$: as nucleons become more tightly bound we can expect the effective nuclear radius to decrease.
However, given that $I^\beta\propto F_{0}(2, E_e) \propto R_0^{-4\alpha^2}$ it is straightforward to see that modification of the Fermi function need not be considered.

\noindent{\bf Phase space integral.}
These points established we can now add the small perturbation $\delta E$ to $E_{\rm max}$ and evaluate the phase space integral for a given choice of Fermi function:
\begin{equation}
    \frac{\delta I^\beta}{I^\beta} = \frac{\int_{m_e}^{E_{\rm max}(0)+\delta E} F_0\left(Z+1, E_e\right) p_e E_e\left(E_{\rm max}(0)+\delta E-E_e\right)^2 \mathrm{~d} E_e}{\int_{m_e}^{E_{\rm max}(0)} F_0\left(Z+1, E_e\right) p_e E_e\left(E_{\rm max}(0)-E_e\right)^2 \mathrm{~d} E_e}-1\,,
\end{equation}
where:
\begin{align}
    F_0\left(Z, E_e\right) = &\left(\frac{\Gamma(3)}{\Gamma(1) \Gamma\left(1+2 \sqrt{1^2-(\alpha)^2}\right)}\right)^2\left(2 p_e R\right)^2\left(\sqrt{1^2-(\alpha)^2-1}\right) \nonumber \\
    &\times\left|\Gamma\left(\sqrt{1^2-(\alpha)^2}+i \alpha \frac{E_e}{p_e}\right)\right|^2 e^{\pi \alpha \frac{E_e}{p_e}}\,,
\end{align}
$R$ is the average radius of a nucleus, a good approximation for the average radius of a nucleus with A nucleons is
\begin{equation}
    R \simeq 1.2\; {\rm fm}\; A^{1 / 3}\\,
\end{equation}
In the Primakoff-Rosen (non-relativistic) approximation\cite{Doi:1985dx}
\begin{align}
     F_{k-1}(Z, E_e) &\simeq F_{k-1}^{\rm{NR}}(Z, E_e) \nonumber \\
     &= F_0^{\rm{PR}}(Z, E_e)[(k-1) !]^{-2} \prod_{j=1}^{k-1}\left[(k-j)^2+y^2\right]\,, 
\end{align}
where $F_{0}^{\rm{PR}}(Z, E_e)=2 \pi y /[1-\exp (-2 \pi \alpha Z)], \; y=\alpha Z E_e / p_e$. We can find the result exactly, expanding for small $\delta E/E_{\rm max}$ to give
\begin{equation}
    \frac{\delta I^\beta}{I^\beta} =\frac{\int_{m_e}^{E_{\rm max}(0)+\delta E} F_0^{\rm{PR}}\left(Z+1, E_e\right) p_e E_e\left(E_{\rm max}(0)+\delta E-E_e\right)^2 \mathrm{~d} E_e}{\int_{m_e}^{E_{\rm max}(0)} F_0^{\rm{PR}}\left(Z+1, E_e\right) p_e E_e\left(E_{\rm max}(0)-E_e\right)^2 \mathrm{~d} E_e}-1\,,
\end{equation}
Using the Primakoff-Rosen approximation, the integral has the analytic expression
\begin{align}
    \frac{\delta I^\beta}{I^\beta} &=\frac{\delta E(5 E_{\rm max}(0)^4+15m_e^4-20E_{\rm max}(0)m_e^3)}{(E_{\rm max}(0)-m_e)^3(6m_e^2+3 E_{\rm max}(0) m_e+E_{\rm max}(0)} \nonumber \\
    &+\frac{\delta E^2(10E_{\rm max}(0)^3-10m_e^3)}{(E_{\rm max}(0)-m_e)^3(6m_e^2+3 E_{\rm max}(0) m_e+E_{\rm max}(0)} \nonumber \\
    &+\frac{\delta E^3 10E_{\rm max}(0)^2}{(E_{\rm max}(0)-m_e)^3(6m_e^2+3 E_{\rm max}(0) m_e+E_{\rm max}(0)} \nonumber \\
    &+\frac{\delta E^4 5E_{\rm max}(0)}{(E_{\rm max}(0)-m_e)^3(6m_e^2+3 E_{\rm max}(0) m_e+E_{\rm max}(0)} \nonumber \\
    &+\frac{\delta E^5}{(E_{\rm max}(0)-m_e)^3(6m_e^2+3 E_{\rm max}(0) m_e+E_{\rm max}(0)}\,,
\end{align}
Expanding around $\delta E=0$ then gives
\begin{align}
    \label{Eq: PR phase space integral}
	\frac{\delta I^\beta}{I^\beta} 
    &\simeq\frac{\delta E(5 E_{\rm max}(0)^4+15m_e^4-20E_{\rm max}(0)m_e^3)}{(E_{\rm max}(0)-m_e)^3(6m_e^2+3 E_{\rm max}(0) m_e+E_{\rm max}(0)} \nonumber \\
    &\simeq 0.162265 \times\left(\frac{\delta E}{\mathrm{keV}}\right), \quad|\delta E| \ll 18.6\, \mathrm{keV}\,,
\end{align}
For the fully relativistic Fermi function given in \eqref{eq: Fermi function}, we have
\begin{equation}
    \frac{\delta I^\beta}{I^\beta} =\frac{\int_{m_e}^{E_{\rm max}(0)+\delta E} F_0\left(Z+1, E_e\right) p_e E_e\left(E_{\rm max}(0)+\delta E-E_e\right)^2 \mathrm{~d} E_e}{\int_{m_e}^{E_{\rm max}(0)} F_0\left(Z+1, E_e\right) p_e E_e\left(E_{\rm max}(0)-E_e\right)^2 \mathrm{~d} E_e}-1\,,
\end{equation}
Numerical integration and expanding around $\delta E=0$ then yields
\begin{equation}
    \label{eq: I}
	\frac{\delta I^\beta}{I^\beta} \simeq 0.184252 \times\left(\frac{\delta E}{\mathrm{keV}}\right), \quad|\delta E| \ll 18.6\, \mathrm{keV}\,,
\end{equation} 
which agrees to within 12\% with the non-relativistic case.

\noindent{\bf Beta strengths.}
Following Ref.~\cite{Faessler:2011zz, Dvornicky:2011fm, Doi:1985dx}, on the basis on isospin symmetry and selection rules it can also be shown that 
\begin{equation}
	\label{Eq: matrix elements}
	M_{F} = 1\,,\quad |M_{GT}|^{2} = 3\,,
\end{equation}
If we assume isospin symmetry, these results should carry over to the present context, as selection rules are presumably not $\theta$-dependent for small values of $\theta$.
The couplings $g_A$ and $g_V$ may also have some $\theta$-dependence, however following Ref.~\cite{Lee:2020tmi} it can also be assumed that their $\theta$-dependence is subleading to phase space modifications.

\noindent{\bf Fractional decay rate.}
These points established we can now combine these results to establish our quantity of interest for comparison to experimental data, the fractional change in the beta decay rate $\Gamma$ as a function of $\theta$,
\begin{equation}
	I_0(\theta)\equiv\frac{\Gamma(\theta)-\Gamma(0)}{\Gamma(0)}
    \simeq\frac{\delta I^\beta(\theta)}{I^\beta(0)}\,,
\end{equation}
where we have used the fact that $\theta$-dependence primarily enters through modification of the phase space integral, and dependence on the various other quantities present in $\Gamma$ cancels.

In the axion dark matter scenario the form of $\theta$ is 
\begin{equation}
    \label{eq: supplementary theta}
	\theta \simeq 2.6\times10^{-4}
    \left(\frac{\rho_{DM}}{0.45\rm{GeV}/\mathrm{cm}^3}\right)^{1/2} 
    \left(\frac{10^{-21}\rm{eV}}{m_a}\right)
    \left(\frac{10^{13}\rm{GeV}}{f_a}\right)    
    \cos(\omega t +\vec p\cdot \vec x+\phi)\,,
\end{equation}
where $\rho_{DM}$ is the DM density, $\omega=m_a(1+v^2+\mathcal{O}(v^4))$, $p$, $v$, $f_a$ and $m_a$ are the axion field momentum, velocity, decay constant and mass respectively, and $\phi$ is an arbitrary phase.
This then yields 
\begin{equation}
    I_0(\theta)\simeq A(\rho, m_a, f_a)\cos(\omega t +\vec p\cdot \vec x+\phi)^2\,,
\end{equation} 
where the amplitude $A$ can easily be found via substitution of equations \eqref{eq: delta theta}, \eqref{eq: I} and \eqref{eq: supplementary theta}.
So, we apparently find a fractional change in the decay rate which is strictly positive relative to the ($\theta=0$) non-axion case.
In simple terms, this would mean that the axion-induced effect apparently only creates an excess in the number of nuclear decays, rather than alternating excesses and deficits.
However, care is required here. 

In a Universe where this mechanism is active the measured values of physical quantities such as decay rates are actually not those at $\theta=0$, but are instead those at $\langle\theta^2\rangle$ (since binding energies are $\mathcal{O}(\theta^2)$ at leading order).
Therefore when dealing with experimental data, rather than comparing $\Gamma(\theta)$ to $\Gamma(0)$, we should instead consider
\begin{equation}
    I(\theta)
    \equiv\frac{\Gamma(\theta)-\langle\Gamma\rangle}{\langle\Gamma\rangle}
    =\frac{\Gamma(\theta)}{\langle\Gamma\rangle} - 1
    =\frac{\Gamma(\theta)}{\Gamma(0)}
    \left(\frac{\langle\Gamma\rangle}{\Gamma(0)}\right)^{-1} - 1\,,
\end{equation}
where $\langle\Gamma\rangle$ is the average value of $\Gamma(\theta)$.
We can calculate $\langle\Gamma\rangle/\Gamma(0)$ by virtue of already knowing $I_0(\theta) = \Gamma(\theta)/\Gamma(0)-1$ and using $\langle\cos(\omega t+\dots)^2\rangle=1/2$.
Since $A<<1$ we then find
\begin{equation}
    \left(\frac{\langle\Gamma(\theta)\rangle}{\Gamma(0)}\right)^{-1}
    \simeq\left(1+A\langle\cos(\omega t +\vec p\cdot \vec x+\phi)^2\rangle\right)^{-1}
    \simeq 1 - \frac{1}{2}A\,.
\end{equation}
Inserting this yields
\begin{equation}
     I(\theta) 
     \simeq \left(1+A\cos(\omega t +\vec p\cdot \vec x+\phi)^2\right)\left(1-\frac{1}{2}A\right)-1
     \simeq A\left(\cos(\omega t +\vec p\cdot \vec x+\phi)^2-\frac{1}{2}\right)\,.
\end{equation}
Using $2\cos(x)^2-1=\cos(2x)$, we then find
\begin{equation}
    I(\theta) 
    \simeq 1.3\times10^{-5}\left(\frac{\rho_{DM}}{0.45\rm{GeV}/\mathrm{cm}^3}\right) \left(\frac{10^{-21}\rm{eV}}{m_a}\right)^2
    \left(\frac{10^{13}\rm{GeV}}{f_a}\right)^2
    \cos(2\omega t +\dots)\,,
\end{equation}
which oscillates about zero, as expected, leading to alternating excesses and deficits in the number of nuclear decays per unit time.
We can also see that the corresponding shift in the decay energy,
\begin{equation}
    \delta E
    \simeq 3.5\times10^{-2}{\rm eV}\left(\frac{\rho_{DM}}{0.45\rm{GeV}/\mathrm{cm}^3}\right) \left(\frac{10^{-21}\rm{eV}}{m_a}\right)^2
    \left(\frac{10^{13}\rm{GeV}}{f_a}\right)^2
    \cos(2\omega t+\dots)\,,
\end{equation}
is also a small correction in regions of the axion parameter space we exclude, justifying our previous assumption that $\delta E<< Q$.
Similarly, $\theta<<1$ in these regions.

\end{document}